\documentclass[aps,physrev]{revtex4-2}
\usepackage{amsmath}
\usepackage{graphicx}
\usepackage{dcolumn}
\usepackage{bm}
\usepackage{subfigure}
\usepackage{color, xcolor}
\usepackage{multirow}

\begin{document}

\title{Optical echoes of light near a black hole}
\author{Suting Ju$^1$}
\author{Jingxuan Zhang$^1$}
\author{Li-Gang Wang$^1$}
\email{lgwang@zju.edu.cn}
\affiliation{$^1$School of Physics, Zhejiang University, Hangzhou 310058, China}
\date{\today}

\begin{abstract}
The light deflection under a strong gravitational field, referred to as strong gravitational lensing, 
provides a powerful probe of spacetime geometry. 
Besides, laboratory analogue models are employed to study the effects of curved spacetime 
and explore the design of optical devices. 
Here, applying the framework of analogue gravity, 
we reveal the behavior of the optical echo from a pulsed point-like source near a black hole, 
which is strongly dependent on the interplay of the black hole's photon sphere and the source's duration. 
We model the Schwarzschild spacetime as a Flamm paraboloid and calculate the echo response, 
using analytical geodesic solutions and the Huygens-Fresnel principle. 
Particularly, when the spatial scale of pulse duration is comparable to the photon sphere, 
continuous ``echo tails" appear along bright interference fringes in temporal response. 
Analysis in both the temporal and frequency domains reveals that these echo tails are a signature of resonance between the incoming pulse and the photon sphere. 
This work provides a wave-optics perspective on the interaction between dynamic sources and black holes, 
offering a table top window on strong gravitational lensing. 
\end{abstract}

\maketitle

\section{Introduction}

General relativity interprets gravity as the geometric effect of spacetime. 
Recently, the horizon-scale observation of the supermassive black holes M87$^{*}$ \cite{EHT2019} 
and Sgr A$^{*}$ \cite{EHT2022} once again proves the validity of the theory. 
The strong gravitational field around the compact celestial object 
will cause severe light deflection, displaying the strong gravitational lensing effect \cite{Tsupko2008,Cunha2018,Johnson2020}. 
Moreover, the strong gravitational lensing allows photons to circle multiple times before escaping to infinity. 
The time delay between two rays orbited different times will cause echoes to the observers. 
The echo response of a transient point light source has been discussed in different four-dimensional spacetimes, 
such as the Schwarzschild spacetime \cite{Zenginoglu2012} and the Kerr spacetime \cite{Wong2021,ACA2024,Wong2024}. 
In the field of time-domain astronomy, the transient lensing as well as its echoes 
are also discussed for the use in detection and application \cite{Liao2022}, 
such as the detection of primordial black holes \cite{Profumo2025}. 
It is evident that, as detection precision continues to improve, 
studying echo signals of a pulsed source near compact objects will be highly promising 
for deepening our understanding and exploration of strong gravitational fields. 

Beyond astronomical observations, analogue gravity offers an alternative
approach to studying the physics of distant celestial bodies. Various physical
systems, including the water tank \cite{Euve2016}, the Bose-Einstein
condensate \cite{Steinhauer2016}, and nonlinear optics experiments 
\cite{Braidotti2022}, have been proposed to mimic black holes and observe
related interesting phenomena. Waveguide-based systems, in particular, have emerged as a
versatile platform for simulating curved spacetime. 
Based on the theory of transformation optics \cite{Plebanski1960,Leonhardt2006,Pendry2006}, 
waveguides with refractive index distributions are designed 
to simulate curved spacetime and study the effects within it, 
such as the celestial mechanics \cite{Genov2009}, the gravitational lensing \cite{Sheng2013}, the Einstein ring \cite{Sheng2016}, the (anti-)de Sitter metric \cite{Chen2020}, and the exterior and interior Schwarzschild metric \cite{Chen2010,Xiao2023}. 
Besides gradient waveguides,
curved-waveguide surfaces can also be utilized to simulate curved space. For
instance, curved waveguides with constant Gaussian curvature, as analogue
models for universes with non-vanishing cosmological constants, have been
used to study the evolution of speckle pattern dynamics \cite%
{Schultheiss2016}. A paraboloid structure inspired by the Schwarzschild
metric was introduced to control light evolution and exhibited a tunneling
effect near its bottleneck \cite{Bekenstein2017}. Curved waveguides based on
the Morris-Thorne transversable wormhole metric were explored using
both flexural waves \cite{Zhu2018} and optical beams \cite{He2020}. 
In fact, there exists the equivalence between gradient waveguides and the curved waveguides \cite{Xu2019,Xu2022,EFG2024}. 
These tabletop experiments bridge the gap between theoretical
relativity and laboratory physics. 
Furthermore, the research about waveguides based analogue gravity opens up new ways 
for light manipulation. For example, the Flamm paraboloid nanostructure allows control of light in many ways \cite{Bekenstein2017}. 
The photon sphere in Schwarzschild spacetime can be introduced to the spatially refractive index distribution 
of a chaotic billiard to realize control of chaos \cite{Xu2023}. 

In this work, inspired by the behavior of light circling multiple geodesics under strong gravitational fields and
prior research on two-dimensional curved waveguides, we focus on multiple
loops of geodesics near the photon sphere on the Flamm
paraboloid, a surface of revolution emerging from Schwarzschild spacetime 
\cite{FlammRepub}. We investigate both the temporal and frequency domain response of a pulsed point-like source, 
building upon the detailed analysis of geodesics on the Flamm paraboloid \cite{Eufrasio2018} 
and the Huygens-Fresnel principle on two-dimensional curved spaces \cite{Xu2021}. 
These results deepen our understanding of echo signals induced by compact celestial objects, 
and can help expand the scope of using curved waveguides to simulate astronomical phenomena, 
promoting the potential application of curved waveguides as optical devices. 

\section{Theory and formula}

Here, we will first explain our theoretical model and the methods of
analysis. As shown in Fig. \ref{fig:FlammFourRegions}(a), we consider the
situation of a light pulse near a Schwarzschild black hole, which can be
mathematically described as a Flamm paraboloid to mimic the spacetime of the
equatorial slice in the vicinity of a black hole. The light source is
randomly placed at the initial point $(r_{i},\varphi _{i})$, and the
observer is located at the point $(r_{{o}},\varphi _{{o}})$. Due to the
rotational symmetry of the surface, without loss of generality, we can
assume that the light source is placed on the line $\varphi _{i}=0$, and
then the observation angle $\varphi _{{o}}$ can be considered to be changed
from $0$ to $\pi $, since the output field is symmetric for $\varphi _{{o}}$
within $(\pi ,2\pi )$. In the below, we briefly present the theoretical
expression of the geodesics on the surface, from which one can calculate the
geodesics and their length, and then we use the Huygens-Fresnel principle 
\cite{Xu2021} to calculate the output field at the observation position.

\subsection{Geodesics on the Flamm paraboloid}

Let us first consider to obtain the geodesics near a black hole. The
spacetime metric for a Schwarzschild black hole is given by \cite{Schwarzschild} 
\begin{equation}
ds^{2}=-(1-\frac{r_{g}}{r})c^{2}dt^{2}+(1-\frac{r_{g}}{r})^{-1}dr^{2}+r^{2}d%
\theta ^{2}+r^{2}\sin ^{2}\theta d\varphi ^{2},
\end{equation}%
where $r_{g}$ represents the radius of the event horizon, and $c$ is the
speed of light. Here we choose a subspace of the Schwarzschild spacetime by
setting $t=const$ and $\theta =\pi /2$ (the equatorial slice), and then the
metric can be reduced into 
\begin{equation}
ds^{2}=\frac{r}{r-r_{g}}dr^{2}+r^{2}d\varphi ^{2}.  \label{eq:FlammMetric}
\end{equation}%
This means the equatorial plane of the Schwarzchild spacetime at a moment.
The metric can also be written in matrix form $g_{\mu \nu },(\mu ,\nu =1,2)$%
, and $g^{\mu \nu }$ is the inverse matrix of $g_{\mu \nu }$. Embedding Eq.~(%
\ref{eq:FlammMetric}) into 3D flat spacetime $ds^{2}=dz^{2}+dr^{2}+r^{2}d%
\varphi ^{2}$, one can get $z=2\sqrt{r_{g}(r-r_{g})}$ as the generatrix for
a surface of revolution that is called the Flamm paraboloid \cite{FlammRepub}.

Light in curved space propagates along its geodesics. On the surface of a
Flamm paraboloid, the geodesics obey the equation 
\begin{equation}
\frac{d^{2}x^{\mu }}{ds^{2}}+\Gamma _{\nu \rho }^{\mu }\frac{dx^{\nu }}{ds}%
\frac{dx^{\rho }}{ds}=0,  \label{eq:GeoEq}
\end{equation}%
with $x^{\mu }=(x^{1},x^{2})=(r,\varphi )$, and the Christoffel symbol $%
\Gamma _{\nu \rho }^{\mu }=\frac{1}{2}g^{\mu \lambda }(\partial _{\nu
}g_{\rho \lambda }+\partial _{\rho }g_{\lambda \nu }-\partial _{\lambda
}g_{\nu \rho })$. The tangent vectors $\frac{dx^{\mu }}{ds}$ of the
geodesics satisfy $g_{\mu \nu }\frac{dx^{\mu }}{ds}\cdot \frac{dx^{\nu }}{ds}%
=1$ according to Eq. (\ref{eq:FlammMetric}). Through analytical calculation,
one can derive the geodesic orbit equation on the surface of Eq. (\ref%
{eq:FlammMetric}) as follow 
\begin{equation}
(\frac{dr}{d\varphi })^{2}=r^{4}(1-\frac{r_{g}}{r})(\frac{1}{r_{p}^{2}}-%
\frac{1}{r^{2}}),  \label{eq:GeoOrbEq}
\end{equation}%
where $r_{p}=r_{i}^{2}(d\varphi /ds)_{i}$ is an integration constant that is
determined by the initial conditions of the geodesic and it can also be seen
as the orbit periastrons for the different geodesics (between the source and
the observer) derived from Eq.~(\ref{eq:GeoEq}). Obviously, when $%
r^{2}=r_{p}^{2}$, the right side of Eq.~(\ref{eq:GeoOrbEq}) goes to zero,
indicating the turning point of the geodesic.

When it comes to the calculation of geodesics on an arbitrary surface, the
most general method is to solve Eq.~(\ref{eq:GeoEq}) numerically. For
numerically solving Eq.~(\ref{eq:GeoEq}), the initial angle will decide
which geodesic is drawn. Since we consider multiple loops of light around
the event horizon on the Flamm paraboloid, finding the corresponding angle is
more difficult. While for the Flamm paraboloid, one can derive not only the
geodesic orbit equation Eq.~(\ref{eq:GeoOrbEq}), but also its analytical
expression. The analytical expressions of the geodesic orbits on the Flamm
paraboloid have been investigated by Eufrasio et al \cite{Eufrasio2018}.
There are two types of geodesic on the Flamm paraboloid, which can be
distinguished by the periastron $r_{p}$ (also corresponding to the impact
parameter according to \cite{Eufrasio2018}). When $r_{p}>r_{g}$, there exist
regular geodesics and all these geodesics will never reach the event
horizon. By integrating Eq.~(\ref{eq:GeoOrbEq}) with respect to 
$r$, from the dimensionless periastron $\widetilde{r}_{p}=r_{p}/r_{g}$
to the dimensionless radius $\widetilde{r}=r/r_{g}$, one can get 
\begin{equation}
\varphi _{r}(\widetilde{r})=2\sqrt{\frac{\widetilde{{r}}{_{p}}}{\widetilde{{r%
}}{_{p}}-1}}\left( F\left[ \frac{\pi }{2},\frac{-2}{\widetilde{{r}}{_{p}}-1}%
\right] -F\left[ \sin ^{-1}\left( \sqrt{\frac{(\widetilde{{r}}{_{p}}-1)(%
\widetilde{r}+\widetilde{{r}}{_{p}})}{2\widetilde{{r}}{_{p}}(\widetilde{r}-1)%
}}\right) ,\frac{-2}{\widetilde{{r}}{_{p}}-1}\right] \right) ,
\label{eq:theta_r_1}
\end{equation}%
where $F$ is the elliptic function of the first kind. The
inverse of Eq.~(\ref{eq:theta_r_1}) is to obtain the regular geodesic as
follow 
\begin{equation}
\widetilde{r}_{r}(\varphi _{r})=\widetilde{{r}}{_{p}}\Big/{\left( 2\mathrm{cn%
}^{2}\left[ \sqrt{\frac{\widetilde{{r}}{_{p}}-1}{4\widetilde{{r}}{_{p}}}}%
\varphi _{r},\frac{-2}{\widetilde{{r}}{_{p}}-1}\right] -1\right) },
\label{eq:rRegular}
\end{equation}%
with $\mathrm{cn}$ the Jacobi ellipse cosine function. The
angle $\varphi _{r}$ varies from $(-\varphi _{r,\infty },\varphi _{r,\infty
})$ with $\varphi _{r,\infty }\equiv \lim_{\widetilde{r}\rightarrow \infty
}\varphi _{r}(\widetilde{r})$ in Eq. (\ref{eq:theta_r_1}). 

When $r_{p}\leq r_{g}$, there exist singular geodesics, which will finally
go to the event horizon. In this case, the lower boundary of
integration in Eq.~(\ref{eq:GeoOrbEq}) becomes $\widetilde{r}_{low}=%
\widetilde{r}_{g}=1$, and the upper boundary is still $\widetilde{r}%
_{upper}=\widetilde{r}$. The expression of such singular geodesics is  
\begin{equation}
\varphi _{s}(\widetilde{r})=2\sqrt{\frac{\widetilde{{r}}{_{p}}}{1-\widetilde{%
{r}}{_{p}}}}\left( F\left[ \frac{\pi }{2},-\frac{1+\widetilde{{r}}{_{p}}}{1-%
\widetilde{{r}}{_{p}}}\right] +iF\left[ i\sinh ^{-1}\left( \sqrt{\frac{(1-%
\widetilde{{r}}{_{p}})(\widetilde{{r}}{_{p}}+\widetilde{r})}{2\widetilde{{r}}%
{_{p}}(\widetilde{r}-1)}}\right) ,\frac{2}{1-\widetilde{{r}}{_{p}}}\right]
\right) ,  \label{eq:theta_r_2}
\end{equation}%
with $\widetilde{r}_{p}<1$, where $\widetilde{r}_{p}$ should be
understood as the relative impact parameter. The inverse of Eq.~(\ref%
{eq:theta_r_2}) is 
\begin{equation}
\widetilde{r}_{s}(\varphi _{s})=\frac{\widetilde{r}_{p}\left( 1-%
\mathrm{sn}^{2}\left[ i\sqrt{\frac{1-\widetilde{r}_{p}}{4\widetilde{r}
_{p}}}\varphi _{s},\frac{2}{1-\widetilde{r}_{p}}\right] \right) }{%
\mathrm{sn}^{2}\left[ i\sqrt{\frac{1-\widetilde{r}_{p}}{4\widetilde{r}%
_{p}}}\varphi _{s},\frac{2}{1-\widetilde{r}_{p}}\right] +\widetilde{r}%
_{p}},  \label{eq:rSingle}
\end{equation}%
with $\mathrm{sn}$ represents the Jacobi ellipse sine function 
and $\varphi _{s}$ varying in the range $(0,\varphi _{{s,\infty }})$ with $\varphi
_{{s,\infty }}=\lim_{\widetilde{r}\rightarrow \infty }\varphi _{s}(%
\widetilde{{r}})$. 

\begin{figure}[tbph]
\includegraphics[width=0.97\textwidth]{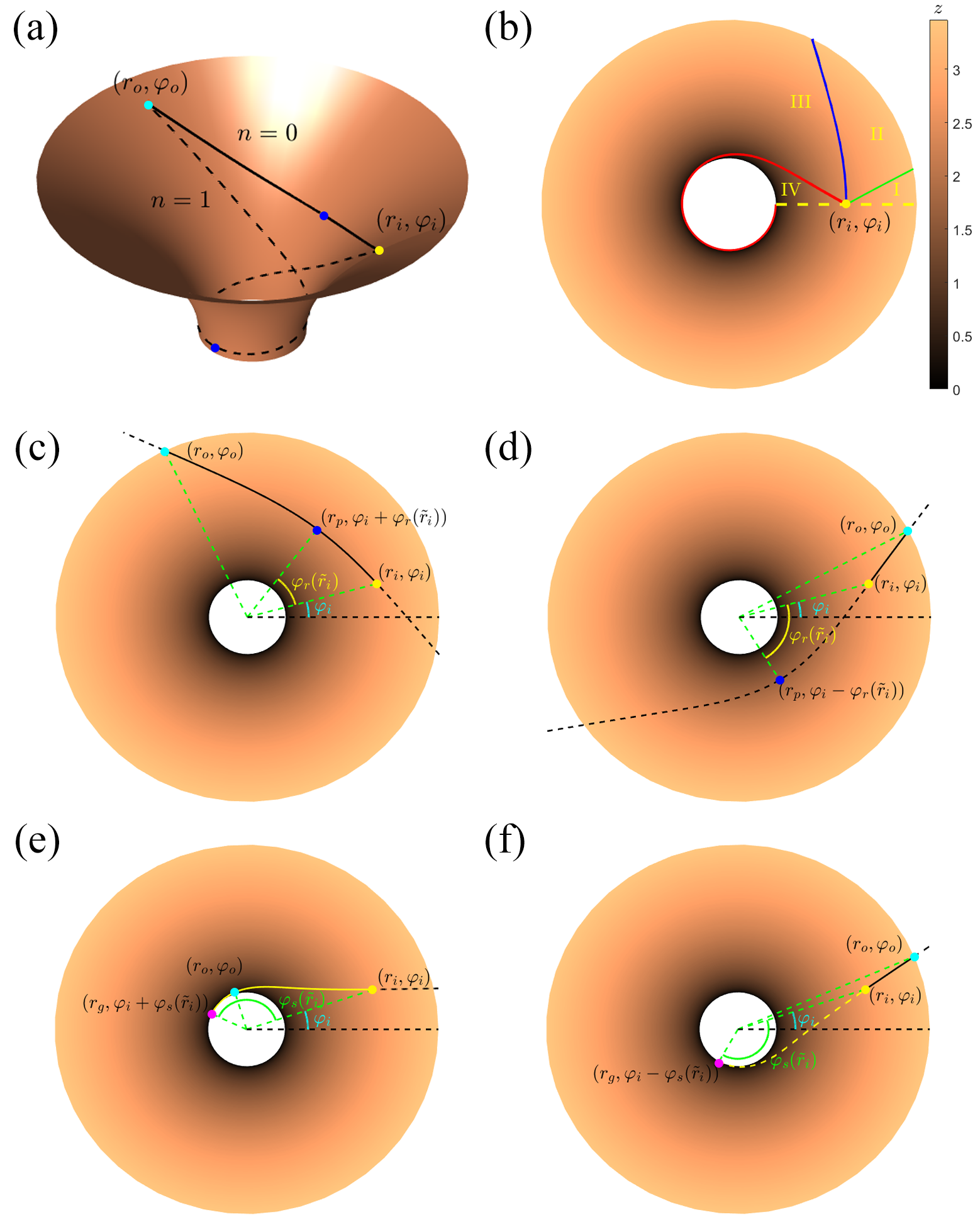}
\caption{Schematic of various geodesics on the Flamm paraboloid. (a) The two
anticlockwise geodesics between the light source and the observer. The
black solid line shows the $n=0$ anticlockwise geodesic, and the black
dotted line shows the $n=1$ anticlockwise geodesic. (b) The diagram of four
regions I, II, III, and IV on the Flamm paraboloid starting from a light source.
The height of the event horizon is set at zero. (c-d) The angular
position of the corresponding periastron for the geodesics located in
regions II and III, and (e-f) the angular position of the intersection of the
corresponding geodesic and the event horizon when the geodesics are in
regions I and IV. Here, the light source is located at $(r_{i},\protect%
\varphi _{i})$ (see the yellow dot), and the observer is $(r_{o},\protect%
\varphi _{o})$ (see the cyan dot), the blue dots mark the periastron of the
geodesics, the magenta dots in (e) and (f) denote the intersection between a
singular geodesic and the event horizon, and the singular geodesic colored
yellow in (e) does not reach $r_{o}$ as (c), (d), and (f).}
\label{fig:FlammFourRegions}
\end{figure}

When the geodesics of light are emitted from the light source located at $%
(r_{i},\varphi _{i})$ which is not coincident with the periastra of those
geodesics, the theoretical expression of geodesic lines in the above need to
be rotated by an additional angle to pass through the light source. As shown
in Fig.~\ref{fig:FlammFourRegions}(b), since light can
propagate in different directions from the source, we separate the Flamm
paraboloid into \textit{four} regions to discuss the anticlockwise geodesic
expression. The green line that divides the regions I and II possesses the
impact parameter $r_{p}=r_{g}$ (i.e., $\widetilde{r}_{p}=1$) , the blue line
that divides the regions II and III possesses the impact parameter $%
r_{p}=r_{i}>r_{g}$ (i.e., $\widetilde{r}_{p}=\widetilde{r_{i}}$, with $%
\widetilde{r_{i}}=r_{i}/r_{g}$), and the red line that divides regions III
and IV also possesses the impact parameter $r_{p}=r_{g}$ (i.e., $\widetilde{r%
}_{p}=1$). In other words, in regions II and III, geodesics are
always regular geodesics with $r_{p}>r_{g}$. While in regions I and IV,
geodesics become singular geodesics. In each region, we need to define the
corresponding rotation angle $\varphi _{j}$, with $j=1,2,3,4$ denoting the
different regions mentioned above. As shown in Fig. \ref%
{fig:FlammFourRegions}(c-f), these angles $\varphi _{j}$, corresponding to
the angular position of the corresponding periastron in regions II and III
or the angular position of the intersection of the corresponding geodesic
and the event horizon in regions I and IV, are given by%
\begin{eqnarray}
{\varphi _{1}} &=&{}\varphi _{i}-\varphi _{s}(\widetilde{r_{i}}),
\label{phai1} \\
{\varphi _{2}} &{=}&\varphi _{i}-\varphi _{r}(\widetilde{r_{i}}),
\label{phai2} \\
{\varphi _{3}} &{=}&\varphi _{i}+\varphi _{r}(\widetilde{r_{i}}),
\label{phai3} \\
\varphi _{4} &=&\varphi _{i}+\varphi _{s}(\widetilde{r_{i}}),  \label{phai4}
\end{eqnarray}%
{\ where }$\widetilde{r_{i}}=r_{i}/r_{g}$.{\ Here, the values of }$\varphi
_{r}(\widetilde{r_{i}})$ and $\varphi _{s}(\widetilde{r_{i}})$ are
calculated from Eqs. (\ref{eq:theta_r_1}) and (\ref{eq:theta_r_2}),
respectively.

For every geodesic between the light source $(r_{i},\varphi _{i})$ and the
observer $(r_{o},\varphi _{o})$, one can use Eq.~(\ref{eq:rRegular}) or Eq.~(%
\ref{eq:rSingle}) to determine its trajectory according to each value of $%
\widetilde{r}_{p}$ (related to both the intial position $r_{i}$ and the
direction of the geodesic starting from the source). Once the angular
position ${\varphi _{j}}$ for the periastron of each geodesic or the
intersection between the geodesic and the event horizon is determined
according to Eqs. (\ref{phai1})-(\ref{phai4}), the relative angular position
of the observer is given by $\delta \varphi _{j}=\varphi _{o}-\varphi _{j}$%
, indicating the angular change from the {periastron or the intersection to
the observer along the corresponding geodesic in different regions. Thus, %
Substituting $(r_{o},\varphi _{o})$ and $\delta \varphi _{j}$ into Eq.~(\ref%
{eq:rRegular}) or Eq.~(\ref{eq:rSingle}), that is to say, setting $%
\widetilde{r}_{r}$ or $\widetilde{{r}}_{s}=\widetilde{r}_{o}$ ($\widetilde{r}%
_{o}=r_{o}/r_{g}$) and $\varphi _{r}$ or $\varphi _{s}=\delta \varphi _{j}$,
one can determine the corresponding geodesic by searching $\widetilde{r}_{p}$%
. To avoid the denominator being zero, Eq.~(\ref{eq:rRegular}) in regions II
and III or Eq.~(\ref{eq:rSingle}) in regions I and IV can be explicitly
written as 
\begin{equation}
2\mathrm{cn}^{2}\left[ \sqrt{\frac{\widetilde{r}_{p}-1}{4\widetilde{r}_{p}}}%
(\varphi _{o}-\varphi _{j}(\widetilde{r}_{p},\widetilde{r}_{i},\varphi
_{i})),\frac{-2}{\widetilde{r}_{p}-1}\right] -1-\frac{\widetilde{r}_{p}}{%
\widetilde{r}_{o}}=0,  \label{eq:EqRegular}
\end{equation}%
and 
\begin{equation}
(\widetilde{r}_{o}+\widetilde{r}_{p})\mathrm{sn}^{2}\left[ i\sqrt{\frac{1-%
\widetilde{r}_{p}}{4\widetilde{r}_{p}}}(\varphi _{o}-\varphi _{j}(\widetilde{%
r}_{p},\widetilde{r}_{i},\varphi _{i})),\frac{2}{1-\widetilde{r}_{p}}\right]
+(\widetilde{r}_{o}-1)\widetilde{r}_{p}=0,  \label{eq:EqSingle}
\end{equation}%
respectively, with $j=1,2,3,4$ in different regions. By solving Eq.~(\ref%
{eq:EqRegular}) or Eq.~(\ref{eq:EqSingle}), the geodesics are determined
through searching the impact parameter $\widetilde{r}_{p}$. If a geodesic
loops around the event horizon for $n$ times, its impact parameter $%
\widetilde{r}_{p}$ can be derived by setting the observation angle as $%
\varphi _{o}\rightarrow \varphi _{o}+2n\pi $ with $n=1,2,3,\cdots $ being
integers, and using the equation Eq.~(\ref{eq:EqRegular}) with $j=3$. In the
process of solving $\widetilde{r}_{p}$, there may be more than one solution
for $n\geq 1$ cases, and the smallest $\widetilde{r}_{p}$ is what we need in
this case after trying.

The geodesics discussed above are anti-clockwise geodesics between the light
source and the observer. For clockwise geodesics, because of the symmetry of
the surface, the impact parameter $\widetilde{r}_{p}$ can be calculated by
setting the angle as $\varphi _{o}\rightarrow 2\pi -\varphi _{o}$.
Therefore, for a given observer $(r_{o},\varphi _{o})$, we can first get two
direct propagation geodesics ($n=0$, anticlockwise and clockwise), then set
the observation angle as $\varphi _{o}\rightarrow \varphi _{o}+2\pi $ to get
two one-loop geodesics $(n=1)$, set the observation angle as $\varphi
_{o}\rightarrow \varphi _{o}+4\pi $ to get two two-loops geodesics $(n=2)$,
and repeat the steps to get more geodesics. The length $L$ of every geodesic
needed for Huygens-Fresnel principle is calculated numerically according to
its geodesic trajectory.

\subsection{Light pulse response}

Now let us to consider how an optical pulse evolves near a black hole. Due
to the strong gravitational spacetime near a black hole, there exist
multi-loop geodesics if a light source is close to a black hole. For
simplicity, here we assume that the initial light pulse is emitted by a
point-like source and thus it can be expressed as 
\begin{equation}
E_{\mathrm{i}}(t)=\mathrm{\exp }(-\frac{t^{2}}{2\tau ^{2}})\mathrm{\exp }%
(i\omega _{0}t),  \label{eq:TimePulseIn}
\end{equation}%
where $\tau $ is the temporal half-width of the pulse, $\omega _{0}=2\pi
f_{0}$ is its central angular frequency with $f_{0}$ being the center
frequency of the pulse. The spectrum of the source is given by
\begin{eqnarray}
\tilde{E}_{\mathrm{i}}(\omega ) &=&\int_{-\infty }^{+\infty }E_{\mathrm{i}%
}(t)\mathrm{\exp }(-i\omega t)dt  \notag \\
&=&\sqrt{2\pi }\tau \exp [-\frac{\tau ^{2}}{2}(\omega -\omega _{0})^{2}].
\label{eq:TimePulseInF}
\end{eqnarray}
Clearly, the source's spectral width $\Delta\omega$ is inversely proportional 
to $\tau$, and it is determined as $\Delta\omega=1/\tau$.

For the sake of simplicity, we first consider the temporal response
contributed by a single geodesic line with length $L$. According to the
Huygens-Fresnel principle on a surface \cite{Xu2021}, multiplying Eq.~(\ref%
{eq:TimePulseInF}) by the propagation factor $\sqrt{\frac{1}{i\lambda }}%
\mathrm{\exp }(-ikL)/L$, the output field in the frequency domain at a
distance $L$ can be written as 
\begin{equation}
\tilde{E}_{o}(\omega ,L)=\frac{\tau }{L}\sqrt{\frac{\omega }{ic}}\exp [-%
\frac{\tau ^{2}}{2}(\omega -\omega _{0})^{2}]\mathrm{\exp }(-ikL),
\label{eq:TimePulseOutF}
\end{equation}%
where $k=\omega /c=2\pi /\lambda $ is the wave number, and $%
\lambda $ is the wavelength. Then the output field in the time domain
can be readily obtained by the inverse Fourier transform of Eq.~(\ref%
{eq:TimePulseOutF}), which is 
\begin{eqnarray}
E_{o}(t,L) &=&\frac{1}{2\pi }\int_{0}^{\infty }\tilde{E}_{o}(\omega ,L)%
\mathrm{\exp }(i\omega t)d\omega  \label{InverseTransform000} \\
&\approx &\frac{1}{L}\sqrt{\frac{\omega _{0}}{i2\pi c}}\mathrm{\exp }[-\frac{%
(t-\frac{L}{c})^{2}}{2\tau ^{2}}]\exp [i\omega _{0}(t-\frac{L}{c})],
\label{eq:TimePulseOut}
\end{eqnarray}%
under the approximation of $\sqrt{\omega }\approx \sqrt{\omega _{0}}$. From
Eq.~(\ref{eq:TimePulseOut}), one can see that the temporal response
contributed by a single geodesic line can also be seen as a Gaussian pulse
(i.e., the shape is unchanged). Compared with the input pulse, the output
Gaussian amplitude part is $\mathrm{\exp }[-(t-\frac{L}{c})^{2}/2\tau ^{2}]$%
, with the time of flight $L/c$ for light to propagate along
the geodesic, and the output phase part is $\mathrm{\exp }[i\omega _{0}(t-%
\frac{L}{c})]$, which is also moved by $L/c$. In fact, Eq.~(\ref%
{InverseTransform000}) can be exactly solved and its exact expression is
given by 
\begin{eqnarray}
E_{o}(t,L) &=&\frac{i}{8L(c\tau \xi )^{1/2}}\exp (-\frac{1}{2}\omega
_{0}^{2}\tau ^{2})\exp (-\frac{\xi ^{2}}{4})  \notag \\
&&\times \left[ \xi ^{2}I_{-\frac{1}{4}}(-\frac{\xi ^{2}}{4})+(\xi ^{2}-2)I_{%
\frac{1}{4}}(-\frac{\xi ^{2}}{4})+\xi ^{2}I_{\frac{3}{4}}(-\frac{\xi ^{2}}{4}%
)+\xi ^{2}I_{\frac{5}{4}}(-\frac{\xi ^{2}}{4})\right] ,
\label{eq:TimePulseOutExact}
\end{eqnarray}%
with $\xi =(t-L/c)/\tau -i\omega _{0}\tau $. Eq.~(\ref{eq:TimePulseOutExact}%
) is the superposition of four modified Bessel functions multiplying a
Gaussian envelope. Our numerical calculation shows that Eq.(\ref%
{eq:TimePulseOut}) is in a good agreement with the exact solution of Eq. (%
\ref{eq:TimePulseOutExact}) unless $\Delta \omega \gg \omega _{0}$.

Since there are multiple geodesics between the observer and the light
source, one needs to include the contributions of the light fields from
multiple geodesics. Here we denote the length of anticlockwise
geodesics looped $n$ times around the black hole as $L_{2n+1}$, and denote
the length of clockwise geodesics looped $n$ times as $L_{2n+2}$. Therefore
the total output field of a light pulse on the Flamm paraboloid is written
as 
\begin{equation}
E_{o,T}(t)=\sum_{n=0}^{\infty }[E_{o}(t,L_{2n+1})+E_{o}(t,L_{2n+2})].
\label{TotalOutPutField}
\end{equation}%
Clearly, the total output fields are the superpositions of the field
components from a series of pairs of anticlockwise and clockwise geodesics,
and their intensities are readily obtained through $I_{o,T}(t)=\left\vert
E_{o,T}(t)\right\vert ^{2}.$ Meanwhile, it is also very interesting to
obtain the spectrum of the total output field on this surface by using the
Fourier transform again

\begin{eqnarray}
\widetilde{E}_{o,T}(\omega ) &=&\int_{-\infty }^{+\infty }E_{o,Total}(t)%
\mathrm{\exp }(-i\omega t)dt  \notag \\
&=&\sum_{n=0}^{\infty }[\tilde{E}_{o}(\omega ,L_{2n+1})+\tilde{E}%
_{o}(\omega ,L_{2n+2})].  \label{OutSpectrum}
\end{eqnarray}%
Therefore, in order to obtain the field dynamic distribution or its spectrum
near the black hole, one needs to calculate the lengths of the anticlockwise
and clockwise geodesics starting from the light source to any position where
the observer may locate at, which have been discussed in the above
subsection.

\section{Results and discussion}

First, we present the results of the geodesic length for all possible
geodesics connecting between the light source and the observer on the Flamm
paraboloid. Table~\ref{tab:GeoLengthValue} shows the length of
some geodesics when the number of loops $n$ is changed from $n=0$ to $n=5$%
. It shows that after the first few loops ($n=0,1,2$), the increasing
length of anticlockwise or clockwise geodesics quickly approaches to the
value of $2\pi r_{g}$. For larger $n$, as $n$ increases to $n+1$, the
larger the number $n$ is, the closer the increment is to $2\pi r_{g}$,
i.e., $\Delta L_{ac}=L_{2n+1}-L_{2n-1}\longrightarrow 2\pi r_{g}$ for
anticlockwise geodesics and $\Delta L_{c}=L_{2n+2}-L_{2n}\longrightarrow
2\pi r_{g}$ for clockwise geodesics. These relationships work well for both
anticlockwise and clockwise cases. 
\begin{table}[tbh]
\caption{\label{tab:GeoLengthValue}
Geodesic length and its increment for both anti-clockwise and
clockwise geodesics under different number of loops $n$ between the source
and an observer. Here the light source is fixed at $(r_{i}/r_{g}=4,\protect%
\varphi _{i}=0)$, and the observer is set at $(\widetilde{r}_o=r_{o}/r_{g}=5,10,20,\protect%
\varphi _{o}=0.25\protect\pi ,0.5\protect\pi ,0.75\protect\pi )$, with the
radius of the event horizon is $r_{g}=1$ cm. $L_{2n+1}$ is the anticlockwise
geodesic length, and $\Delta L_{ac}=L_{2n+1}-L_{2n-1}$ is its increment.
Similarly, $L_{2n+2}$ is the clockwise geodesic length, and $\Delta
L_{c}=L_{2n+2}-L_{2n}$ is its increment. }%
\begin{ruledtabular}
\begin{tabular}{ccccccccc}
$(\widetilde{r}_o,\varphi_o)$ & $n$ & 0 & 1 & 2 & 3 & 4 & 5\\
\colrule
\multirow{4}{*}{(5,$\pi/4$)} & $L_{2n+1}/r_g$ & 3.6260 & 15.3770 & 21.6983 & 27.9819 & 34.2651 & 40.5482\\
& $\Delta L_{ac}/r_g$ & - & 11.7511 & 6.3212 & 6.2836 & 6.2832 & 6.2832\\
& $L_{2n+2}/r_g$ & 13.7200 & 20.1253 & 26.4098 & 32.6930 & 38.9762 & 45.2594\\
& $\Delta L_{c}/r_g$ & - & 6.4053 & 6.2845 & 6.2832 & 6.2832 & 6.2832\\
\colrule
\multirow{4}{*}{(5,$\pi/2$)} & $L_{2n+1}/r_g$ & 6.5922 & 16.1794 & 22.4841 & 28.7676 & 35.0508 & 41.3339\\
& $\Delta L_{ac}/r_g$ & - & 9.5872 & 6.3047 & 6.2834 & 6.2832 & 6.2832\\
& $L_{2n+2}/r_g$ & 12.8303 & 19.3386 & 25.6241 & 31.9073 & 38.1905 & 44.4737\\
& $\Delta L_{c}/r_g$ & - & 6.5083 & 6.2855 & 6.2832 & 6.2832 & 6.2832\\
\colrule
\multirow{4}{*}{(5,$3\pi/4$)} & $L_{2n+1}/r_g$ & 8.9452 & 16.9729 & 23.2684 & 29.5517 & 35.8349 &42.1181\\
& $\Delta L_{ac}/r_g$ & - & 8.0277 & 6.2955 & 6.2833 & 6.2832 & 6.2832\\
& $L_{2n+2}/r_g$ & 11.8407 & 18.5527 & 24.8399 & 31.1232 & 37.4063 & 43.6895\\
& $\Delta L_{c}/r_g$ & - & 6.7120 & 6.2872 & 6.2832 & 6.2832 & 6.2832\\
\colrule

\multirow{4}{*}{(10,$\pi/4$)} & $L_{2n+1}/r_g$ & 8.1197 & 20.7087 & 27.0331 & 33.3168 & 39.6000 & 45.8832\\
& $\Delta L_{ac}/r_g$ & - & 12.5890 & 6.3244 & 6.2837 & 6.2832 & 6.2832\\
& $L_{2n+2}/r_g$ & 19.0435 & 25.4601 & 31.7448 & 38.0280 & 44.3111 & 50.5943\\
& $\Delta L_{c}/r_g$ & - & 6.4166 & 6.2846 & 6.2832 & 6.2832 & 6.2832\\
\colrule
\multirow{4}{*}{(10,$\pi/2$)} & $L_{2n+1}/r_g$ & 11.1736 & 21.5125 & 27.8191 & 34.1025 & 40.3857 & 46.6689\\
& $\Delta L_{ac}/r_g$ & - & 10.3389 & 6.3065 & 6.2835 & 6.2832 & 6.2832\\
& $L_{2n+2}/r_g$ & 18.1421 & 24.6733 & 30.9590 & 37.2422 & 43.5254 & 49.8086\\
& $\Delta L_{c}/r_g$ & - & 6.5312 & 6.2857 & 6.2832 & 6.2832 & 6.2832\\
\colrule
\multirow{4}{*}{(10,$3\pi/4$)} & $L_{2n+1}/r_g$ & 13.9052 & 22.3068 & 28.6033 & 34.8867 & 41.1698 & 47.4530\\
& $\Delta L_{ac}/r_g$ & - & 8.4016 & 6.2965 & 6.2833 & 6.2832 & 6.2832\\
& $L_{2n+2}/r_g$ & 17.1232 & 23.8873 & 30.1749 & 36.4581 & 42.7413 & 49.0245\\
& $\Delta L_{c}/r_g$ & - & 6.7641 & 6.2875 & 6.2832 & 6.2832 & 6.2832\\
\colrule

\multirow{4}{*}{(20,$\pi/4$)} & $L_{2n+1}/r_g$ & 18.1999 & 31.0476 & 37.3737 & 43.6573 & 49.9405 & 56.2237\\
& $\Delta L_{ac}/r_g$ & - & 12.8477 & 6.3261 & 6.2837 & 6.2832 & 6.2832\\
& $L_{2n+2}/r_g$ & 29.3782 & 35.8006 & 42.0853 & 48.3685 & 54.6517 & 60.9349\\
& $\Delta L_{c}/r_g$ & - & 6.4224 & 6.2847 & 6.2832 & 6.2832 & 6.2832\\
\colrule
\multirow{4}{*}{(20,$\pi/2$)} & $L_{2n+1}/r_g$ & 21.1234 & 31.8521 & 38.1596 & 44.4430 & 50.7262 & 57.0094\\
& $\Delta L_{ac}/r_g$ & - & 10.7287 & 6.3074 & 6.2835 & 6.2832 & 6.2832\\
& $L_{2n+2}/r_g$ & 28.4706 & 35.0138 & 41.2996 & 47.5828 & 53.8659 & 60.1491\\
& $\Delta L_{c}/r_g$ & - & 6.5432 & 6.2858 & 6.2832 & 6.2832 & 6.2832\\
\colrule
\multirow{4}{*}{(20,$3\pi/4$)} & $L_{2n+1}/r_g$ & 24.0198 & 32.6469 & 38.9439 & 45.2272 & 51.5104 & 57.7936\\
& $\Delta L_{ac}/r_g$ & - & 8.6270 & 6.2970 & 6.2833 & 6.2832 & 6.2832\\
& $L_{2n+2}/r_g$ & 27.4354 & 34.2277 & 40.5154 & 46.7986 & 53.0818 & 59.3650\\
& $\Delta L_{c}/r_g$ & - & 6.7923 & 6.2877 & 6.2832 & 6.2832 & 6.2832\\
\end{tabular}
\end{ruledtabular}
\end{table}

For a light pulse, it takes more time to pass through longer geodesics, so
the observer may receive a series of pulses. Here we use the incremental
value of two consecutive anticlockwise or clockwise geodesic lines to define
the echo time that may be observed near a black hole as follow 
\begin{equation}
\tau _{\mathrm{echo}}=2\pi r_{g}/c,
\end{equation}%
which will be manifested in the time evolution of pulse response. 
In our calculation, we numerically calculated the geodesics from 
$n=0$ to $n=4$, and used the $2\pi r_{g}$ relationship for the larger $n$
cases. 

\begin{figure}[tbph]
\includegraphics[width=1\textwidth]{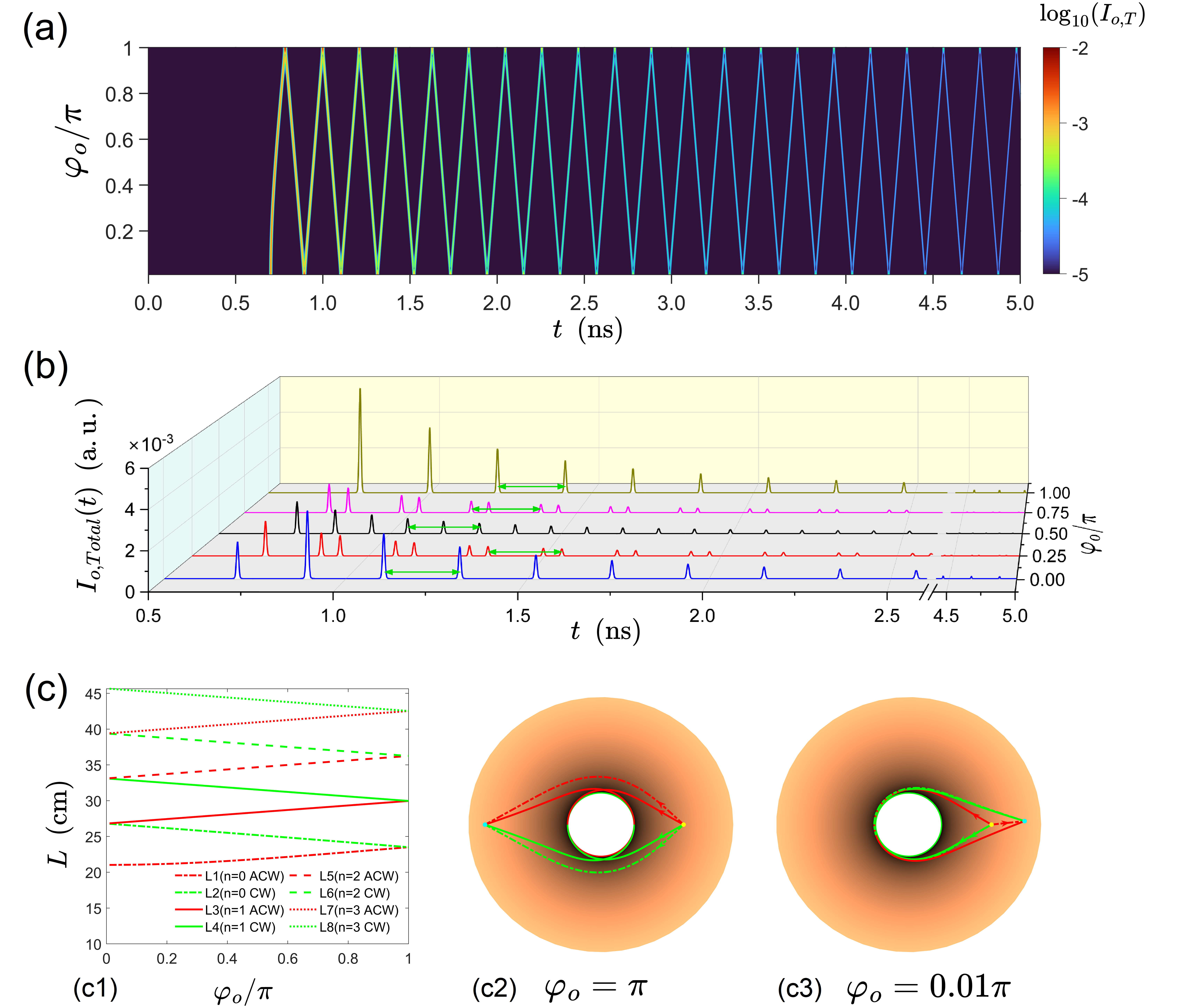}
\caption{Temporal response of a point-like pulsed source on the Flamm paraboloid. 
(a) Temporal response with different observation angles. 
The horizontal axis represents the time $t$ varying from 0 to 5 ns, 
and the vertical axis represents the observation angle 
$\varphi _{o}$ varying from $0.01\pi$ to $\pi $. The color shows
the logarithm of the output intensity $I_{o,T}=\left | E_{o,T} \right | ^2$. 
(b) Five transverse slices of (a) at different observation angles 
$\varphi _{o}=0.01\pi$ (blue line), $\pi /4$ (red line), $\pi /2$ (black line), 
$3\pi /4$ (magenta line), and $\pi $ (dark yellow line). 
The echo time ($\tau _{\mathrm{echo}}=0.21$ ns) is marked with green
double arrows. 
(c) Geodesic length with different observation angles. 
(c1) Red lines show the anti-clockwise geodesics and green lines show the
clockwise geodesics. Dot-dash lines, solid lines, dash lines, and dot lines
show $n=0$, $n=1$, $n=2$, $n=3$ geodesics, respectively. 
(c2) The schematic of the symmetry between $n=0$ anti-clockwise geodesic 
(red dot-dash line) and $n=0$ clockwise geodesic (green dot-dash line). 
(c3) The schematic of the symmetry between $n=0$ clockwise geodesic 
(green dot-dash line) and $n=1$ anti-clockwise geodesic (red solid line).
Other parameters are the radius of event horizon $r_{g}=1$ cm, 
the position of light source $(r_{i}/r_g=1.1, \varphi_{i}=0)$, 
the radius of the observer $r_{o}/r_g=20$, center frequency 
$f_0=\omega_0/2\pi=30$ GHz, the initial
half-width $\tau=0.003$ ns, and the light speed $c=30$ cm$\cdot $GHz. }
\label{fig:TimeDomain}
\end{figure}

As shown in Fig.~\ref{fig:TimeDomain}, the temporal response is made up of a
primary pulse and infinitely many echo pulses. The initial half-width $\tau$
is set much smaller than the echo time $\tau _{\mathrm{echo}}$. Fig.~\ref%
{fig:TimeDomain}(a) shows the output temporal response with observation angle
varying from 0 to $\pi $. It can be seen from the figure that the light
pulses experience a process from separation to merger. 
In the region of backward scattering, the echo pulses gradually 
separate into two pulses. While in the region of forward scattering, 
the light pulses gradually merge. Fig.~\ref{fig:TimeDomain}%
(b) show five slices in Fig.~\ref{fig:TimeDomain}(a). 
As the output angle changes from 0 to $\pi/4$, we can clearly see the 
separation of pulses. Except for the first pulse, all the echo pulses
separate into two pulses, and the intensity of the echo pulses declines. 
When the output angle changes to $\pi/2$, 
where the backward scattering transforms to the forward scattering,
the light pulses become fully separated, 
and the number of pulses is twice of those at $\varphi_o=0$ and $\pi$. 
When the output angle changes from $3\pi/4$ to $\pi$, 
the pulses start to merge. For instance, the second pulse gets close to the
primary pulse and gets away from the third pulse, while the third pulse gets
close to the fourth pulse. When the output angle changes to $\pi$, the merger 
completes and the intensity of the echo pulses increases. The intensity of
the echo pulses has the relationship of $I\propto 1/t^{2}$, which can be
derived from Eq.~(\ref{eq:TimePulseOut}) by substituting $t=L/c$. 

The angular-dependent temporal response arises from 
the variation in geodesic paths with observation angle. 
Fig.~\ref{fig:TimeDomain}(c1) shows
the length of 8 geodesics at different observation angles $\varphi_{o}$. 
In the process of $\varphi_{o}$ changing from 0 to $\pi$, 
the length of anti-clockwise geodesics keeps increasing, and the length of clockwise geodesics
keeps declining. Each line in Fig.~\ref{fig:TimeDomain}(a) is actually contributed by a
geodesic, and the primary pulse is contributed by the $n=0$ anti-clockwise
geodesic. When $\varphi_{o}=\pi$, the anti-clockwise
geodesics and clockwise geodesics with the same number $n$ have the same
length, and when $\varphi_{o}=0$, the $n$-th clockwise geodesic have the same length with $%
(n+1)$-th anti-clockwise geodesic as the schematics shown in Fig.~\ref%
{fig:TimeDomain}(c2) and (c3), indicating the symmetry of the geodesics on the surface. 
Therefore, the separation and merger of echo pulses manifest the 
change of geodesic length. The separation of temporal pulses signifies that
the difference between the $n$-th clockwise geodesic and the $(n+1)$-th
anti-clockwise geodesic gradually increases. Conversely, the merging of
temporal pulses indicates that the difference between the clockwise and
anti-clockwise geodesics with number $n$ gradually decreases. 

\begin{figure}[tbph]
\includegraphics[width=1\textwidth]{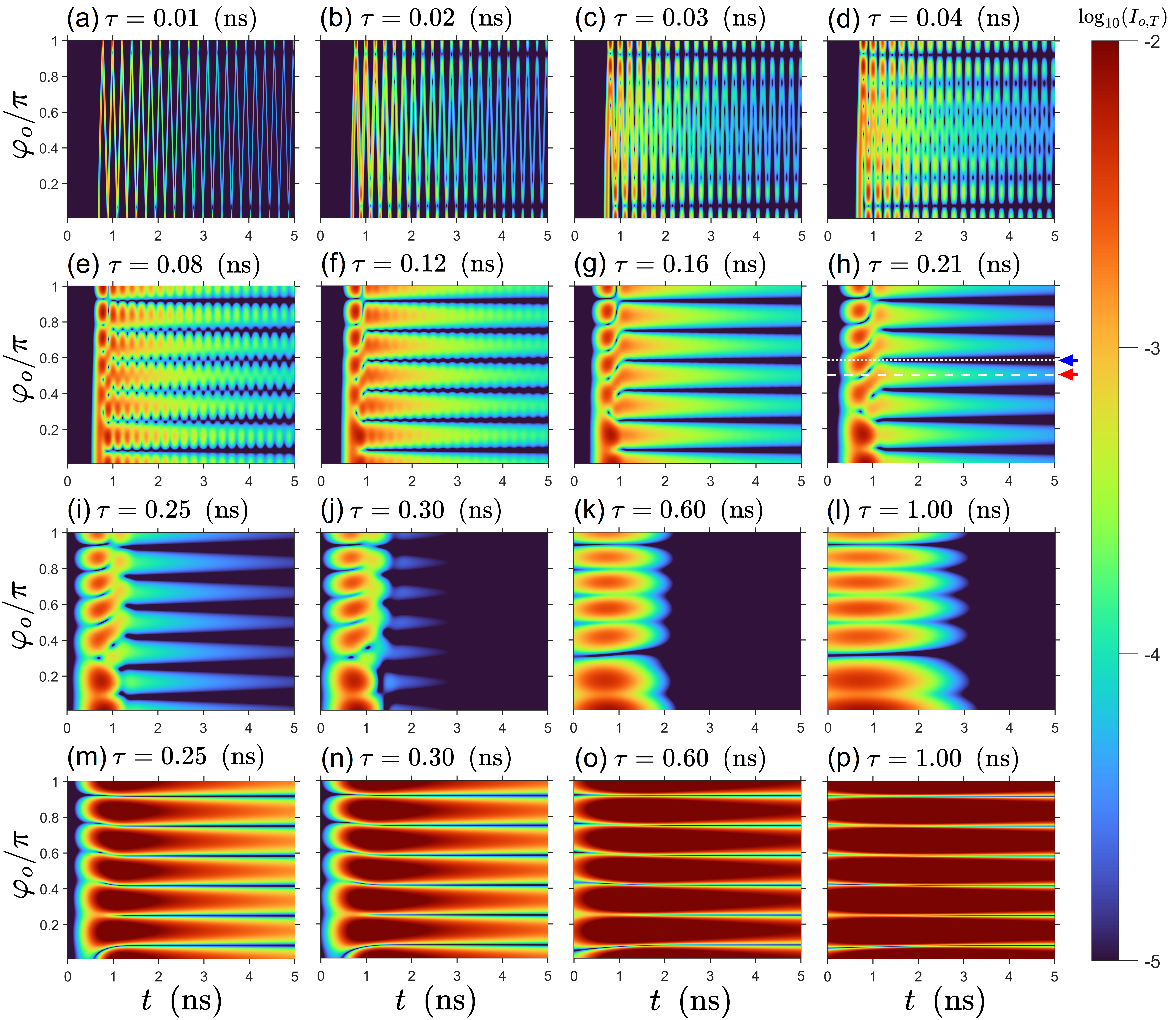}
\caption{
Effect of pulse duration on the evolution of the output fields at
the observer. In (a)-(l), the pulse half-widths are $\tau =0.01$ ns, $0.02$ ns, $%
0.03$ ns, $0.04$ ns, $0.08$ ns, $0.12$ ns, $0.16$ ns, $0.21$ ns, $0.25$ ns, $0.30$%
ns, $0.60$ ns, and $1.00$ ns, respectively, with the center frequency $f_0=30$ GHz. 
In (m)-(p), the pulse widths are the same with (i)-(l), 
but the center frequency is shifted to $f_0=28.65$ GHz. 
Other parameters are the same as in Fig.~\ref{fig:TimeDomain}. 
}
\label{fig:TimeDomain_different_tau}
\end{figure}
When the initial half-width $\tau$ increases, 
the temporal response will change accordingly. 
As the pulse width increases from Fig.~\ref{fig:TimeDomain_different_tau}(a) to (d), 
the angular dark fringes start to occur, but we can still distinguish the echo pulses. 
From Fig.~\ref{fig:TimeDomain_different_tau}(e) to (h), 
as the initial half-width $\tau$ getting close to the echo time $\tau _{\mathrm{echo}}$, 
the angular interference pattern become more obvious. 
Along the bright fringe 
(see the white dash line marked with the red arrow in Fig.~\ref{fig:TimeDomain_different_tau}(h)), 
the enhanced-intensity echo signals form the long ``echo tail" after the primary pulse, 
indicating the temporal structure of the pulse is coupled with the photon sphere. 
While along the dark fringe 
(see the white dot line marked with the blue arrow in Fig.~\ref{fig:TimeDomain_different_tau}(h)), 
the echo pulses cancelled each other, so there is only a single pulse in temporal domain. 
The interference fringes are caused by coherent superposition between adjacent pulses, 
which is actually the result of the geodesic length difference at different observation angles. 
From Fig.~\ref{fig:TimeDomain_different_tau}(i) to (l), 
the echo tails gradually disappear and the width of the primary pulse further increases 
simply as $\tau$ increases. 
However, if we choose the appropriate center frequency $f_0$, 
the echo tails will not disappear but grow stronger as the pulse width increases 
as shown in Fig.~\ref{fig:TimeDomain_different_tau}(m) to (p). 
This effect means the continuous coupling of light pulses with the photon sphere. 
We will next explain this behavior in the frequency domain. 

\begin{figure}
\includegraphics[width=1\textwidth]{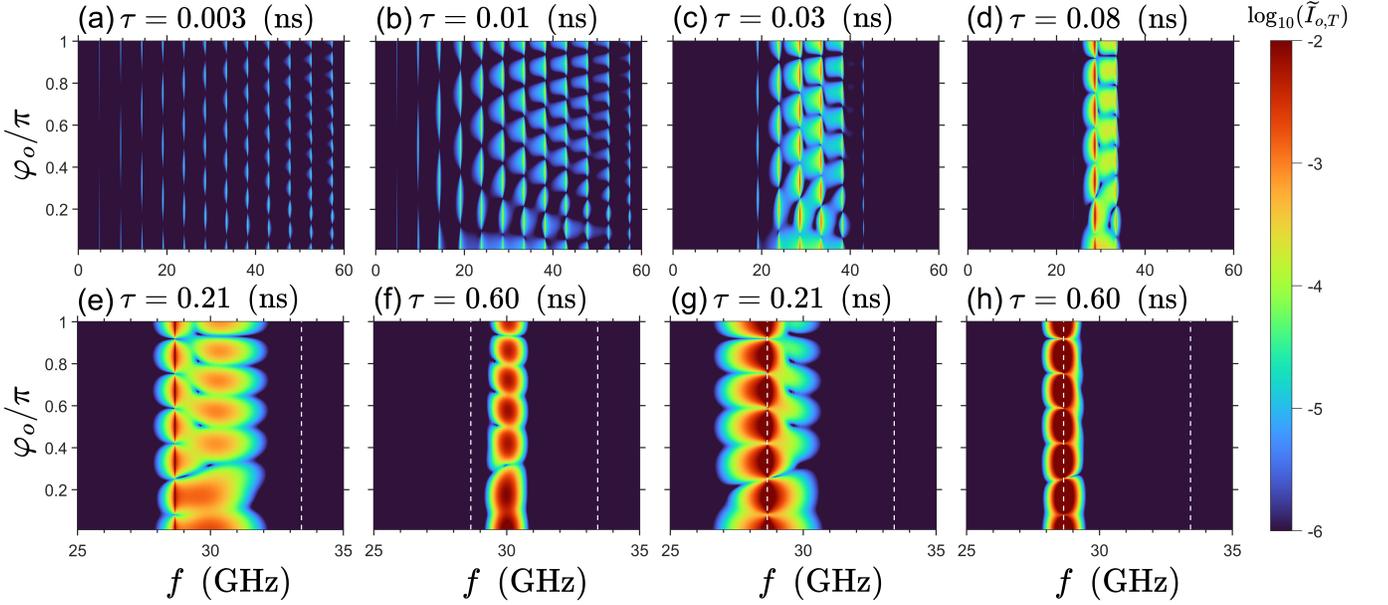}
\caption{
The effect of pulse duration on the frequency domain response. 
The result is derived by $\widetilde{I}_{o,T}(\omega)=\left | \widetilde{E}_{o,T}(\omega) \right |^2 $. 
In (a)-(d), the pulse widths are $\tau=$0.003 ns, 0.01 ns, 0.03 ns, and 0.08 ns, 
with $f=\omega/2\pi$ ranging from 0 to 60 GHz. 
In (e)-(f), the pulse widths are $\tau=$0.21 ns, and 0.60 ns, 
with $f=\omega/2\pi$ ranging from 25 to 35 GHz. 
(g)-(h) are plotted with the same pulse width range as (e)-(f), 
but for a different incident center frequency. 
In (a)-(f), the incident center frequency is $f_0=30$ GHz consistent with those in Fig.~\ref{fig:TimeDomain_different_tau}(a) to (l), 
while in (g)-(h) the incident center frequency is $f_0=6f_{\rm{echo}}=28.65$ GHz 
consistent with those in Fig.~\ref{fig:TimeDomain_different_tau}(m) to (p). 
}
\label{fig:FreqDomain}
\end{figure}
We then investigate the pulse response in the frequency domain. According
to the echo time $\tau_{\mathrm{echo}}$, we can derive the corresponding
fundamental echo frequency 
\begin{equation}
f_{\mathrm{echo}}=1/\tau_{\mathrm{echo}}=c/2\pi r_{\mathrm{g}}.
\end{equation}
For light whose frequency equals any integer multiple of fundamental echo frequency, 
its energy will be trapped by the photon sphere. 
In the case of Fig.~\ref{fig:FreqDomain}, the echo frequency is $f_{\mathrm{echo}}=4.77$ GHz for $r_g=1$ cm. 
In Fig.~\ref{fig:FreqDomain}(a), 
the discrete frequencies are all integer multiples of the fundamental echo frequency. 
These frequency components represent the behavior of the geodesic circulation, 
and in fact represent the photon sphere modes (see Appendix A). 
From Fig.~\ref{fig:FreqDomain}(a) to (d), as the pulse width becomes widen, 
the frequency distribution becomes more concentrated around the center frequency $f_0$, 
and the number of discrete frequencies trapped by the photon sphere reduces. 
The change of frequency distribution is consistent with the variation in the time domain. 
For example, in Fig.~\ref{fig:FreqDomain}(b), there are many photon sphere modes in the frequency domain. 
These frequency components superimpose incoherently, 
forming the angular discontinuous delay-time distribution in Fig.~\ref{fig:TimeDomain_different_tau}(a). 
In Fig.~\ref{fig:FreqDomain}(d), there are two photon sphere modes in the frequency domain, and
their superposition produces a beat pattern in the time domain (see Fig.~\ref{fig:TimeDomain_different_tau}(e)). 
In Fig.~\ref{fig:FreqDomain}(e), the only photon sphere mode 
corresponds to the echo tails in Fig.~\ref{fig:TimeDomain_different_tau}(h). 
When the center frequency is not an integer multiple of the fundamental echo frequency, 
as the pulse width further increases, the frequency components of photon sphere disappear. 
The remaining frequency distribution in Fig.~\ref{fig:FreqDomain}(f) 
is contributed by geodesics without the looping behavior, 
which is actually the $n=0$ anti-clockwise and clockwise geodesics. 
In the corresponding time domain (see Fig.~\ref{fig:TimeDomain_different_tau}(k)), 
the echo tails also fade away. 
When the center frequency is set on one of the photon sphere mode 
(see Fig.~\ref{fig:FreqDomain}(g) and (h)), 
there is always an echo component present in the frequency domain, 
and thus the echo tails in the time domain do not disappear but are amplified
just as Fig.~\ref{fig:TimeDomain_different_tau}(m) to (p) show. 
In other words, each photon sphere mode in the frequency domain 
constitutes corresponding echo tails in the time domain. 
When there are multiple photon sphere modes in the frequency domain, 
the echo tails overlap with each other, 
forming the delay-time distribution. 
Each photon sphere mode appears in the frequency domain as a vertical line 
broken at some specific angles, indicating nodes of zero intensity. 
For a certain photon sphere mode with $M$ times the fundamental echo frequency, 
points with zero intensity occur at observation angles $\varphi_o=(m+1/2)\pi/M$, $m=0,\cdots,M-1$. 
These positions correspond to the dark fringes in the time domain and the number of fringes is also $M$, 
which can be seen in the next part. 

\begin{figure}[tbph]
\includegraphics[width=0.5\textwidth]{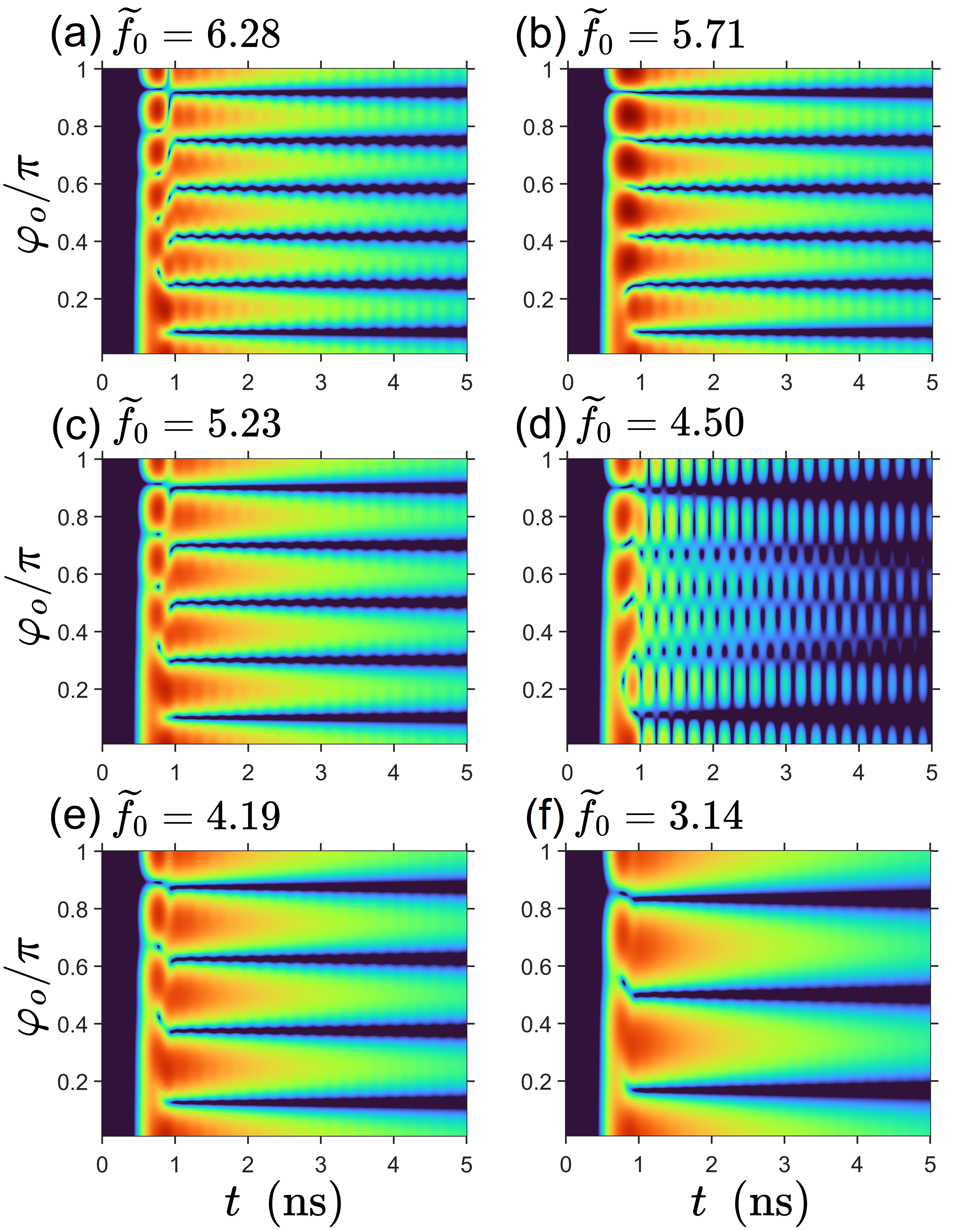}
\caption{
Effect of the center frequency $f_0$ on the evolution of the output fields at the observer. 
In (a)-(f), the values of $\widetilde{f}_0=f_0/f_{\rm{echo}}$ are 6.28, 5.71, 5.23, 4.50, 4.19, and 3.14, respectively. 
The pulse half-width is set as $\tau=0.12$ ns.
Other parameters are the same as in Fig.~\ref{fig:TimeDomain}.
}
\label{fig:TimeDomain_different_lambda0}
\end{figure}
Next, we show how the center frequency $f_0$ influences 
the number of the angular interference fringes in the temporal response. 
As it can be seen from Fig.~\ref{fig:TimeDomain_different_lambda0}(a) 
to Fig.~\ref{fig:TimeDomain_different_lambda0}(f), when the incident center frequency decreases, 
the fringe number in the time domain response decreases accordingly. 
These stripes distribute evenly within the range of 0 to $\pi$. 
Here we use the expression $\widetilde{f}_0=f_0/f_{\rm{echo}}=2\pi r_g/\lambda_0$ to explain 
the fringe number, which is also the resonance condition expression of the ring resonator \cite{McCall1992,Wang2023}. 
The number of dark fringes can be given by $M=\left \lfloor \widetilde{f}_0+1/2 \right \rfloor$. 
Fig.~\ref{fig:TimeDomain_different_lambda0}(a) and (b) both exhibit six dark stripes, 
where the $M=6$ photon sphere mode is the closest in frequency to their respective center frequencies. 
The angular positions of dark fringes are $\varphi_o=(m+1/2)\pi/6$, $m=0,\cdots,5$, 
and the bright fringe positions are $\varphi_o=m\pi/6$, $m=0,1,\cdots,6$. 
In Fig.~\ref{fig:TimeDomain_different_lambda0}(c), 
the center frequency shifts towards the $M=5$ mode, reducing the fringe number to five. 
Therefore, the photon sphere mode closest to the center frequency has the highest intensity in the frequency domain, 
and thus the corresponding interference pattern dominates in the time domain. 
In Fig.~\ref{fig:TimeDomain_different_lambda0}(d), the delay-time distribution is 
contributed by $M=5$ photon sphere mode and the $M=4$ photon sphere mode equally, 
whose echo tails interfere with each other, resulting in interlaced fringes (see Appendix B). 
In Fig.~\ref{fig:TimeDomain_different_lambda0}(e) and (f), 
there exist four and three dark fringes in the time domain, 
which are governed by the $M=4$ and $M=3$ photon sphere mode, respectively. 
The photon sphere modes arising from multi-loops of geodesics 
makes the angular delay-time distribution on the Flamm paraboloid 
similar to the resonances of the ring cavity \cite{Vahala2003}. 

\begin{figure}[tbph]
\includegraphics[width=0.85\textwidth]{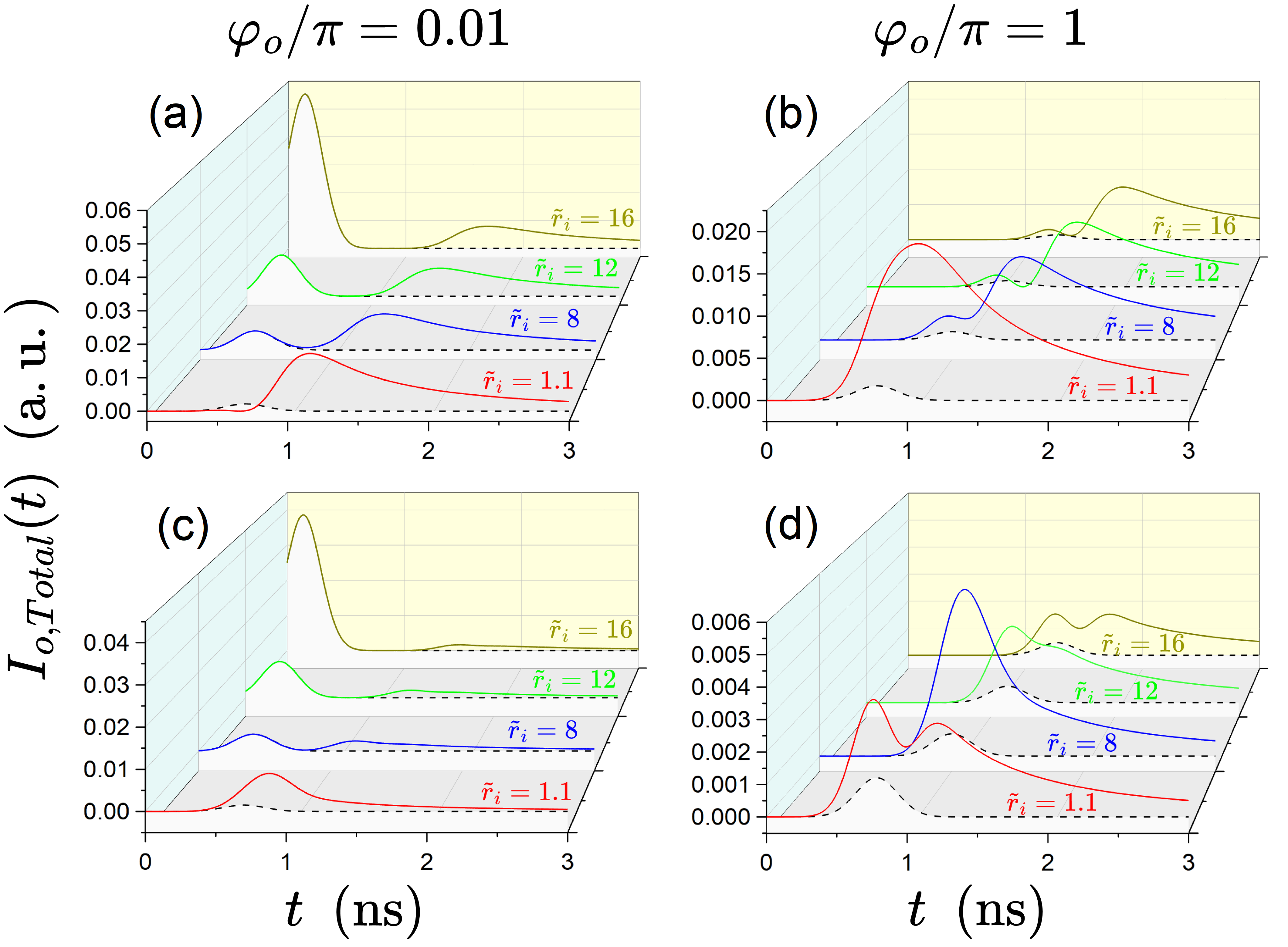}
\caption{
Effect of the source position on the intensity distribution. 
The source positions $\widetilde{r}_i=r_i/r_g=1.1,8,12,16$ are denoted in red, blue, green, and dark yellow lines,  respectively. 
The black dash lines are the pulses contributed by the n=0 anti-clockwise geodesic. 
The observation angle is $\varphi_o/\pi=0.01$ for (a) and (c), 
and $\varphi_o/\pi=1$ for (b) and (d). 
Their center wave length is $\lambda_0/r_g=2\pi/6\approx 1.05$ (i.e. $\widetilde{f}_0=6$) for (a) and (b), 
and $\lambda_0/r_g=1.5$ (i.e. $\widetilde{f}_0=4.19$) for (c) and (d). 
The pulse half-width is $\tau=0.21$ ns. 
Other parameters are the same as in Fig.~\ref{fig:TimeDomain}. 
}
\label{fig:TimeDomain_different_r0}
\end{figure}
The variation introduced by the source position in the temporal response 
mainly occurs in the vicinity of the primary pulse. 
Here we choose the pulse half-width $\tau=0.21$ ns, 
so that there is only one photon sphere mode left in the frequency domain. 
In the direction of $\varphi_o/\pi=0.01$ (see Fig.~\ref{fig:TimeDomain_different_r0}(a) and (c)), 
the observer and the light source are on the same side of the black hole. 
The output line of intensity varying with time consists of two parts, 
which are the primary pulse contributed by the $n=0$ anti-clockwise geodesic 
and the echo tail contributed by other geodesics. 
When the light source is close to the black hole ($\widetilde{r}_i=1.1$), 
the primary pulse and the echo tail are also close to each other, 
forming a single peak in the time domain. 
As the source moves farther away ($\widetilde{r}_i=8,12,16$), 
the time interval between the primary pulse and the echo tail widens and two peaks appear. 
The peak intensity of the primary pulse increases, 
while the echo tails remain almost the same, because there is only one photon sphere mode excited. 
In terms of geodesics, the separation between the primary pulse and the echo tail 
in backward scattering corresponds to the path length difference $(L_2-L_1)/c$ 
between the $n=1$ clockwise geodesic and the $n=0$ anti-clockwise geodesic. 
The higher-order geodesic differences and their corresponding temporal responses 
are insensitive to the changes in the position of the light source. 
When the source position increases, the length of n=1 anti-clockwise geodesic $L_1$ decreases, 
and the length difference $(L_2-L_1)/c$ increases, 
resulting in a greater peak intensity of the primary pulse and a greater distance from the echo tail. 
In the direction of $\varphi_o/\pi=1$ (see Fig.~\ref{fig:TimeDomain_different_r0}(b) and (d)), 
the observer and the light source are on the opposite side of the black hole. 
The output line would appear one peak or two peaks with different source positions. 
Due to the symmetry of the curved surface, 
the anti-clockwise and clockwise geodesics have equal lengths, 
so the peaks of the intensity line are mainly decided by the difference of 
$n=1$ anti-clockwise geodesic and $n=0$ anti-clockwise geodesic $(L_3-L_1)/\lambda_0$. 
When the value of $(L_3-L_1)/\lambda_0$ is close to half-integers, 
the pulses contributed by the $n=1$ anti-clockwise geodesic and $n=2$ anti-clockwise geodesic 
cancel each other, and two peaks appear after superposition. 
On the contrary, when the value of $(L_3-L_1)/\lambda_0$ is close to an integer, 
a single peak will appear after the superposition (see Appendix C). 

\section{Conclusion}

In summary, we have investigated the optical echoes of a pulse point source 
near a black hole in a subspace of the Schwarzschild spacetime. 
Based on the analytical multi-looped geodesics and Huygens-Fresnel principle, 
the response expression of a pulsed source propagating around the black hole has been derived. 
Due to the existence of the photon sphere surrounding the black hole, 
the output pulse response at the observation plane is made up of a series of pulses, 
consisting of a primary pulse and multiple echo pulses. 
We demonstrate a correspondence between the echo tails in the time domain 
and the photon sphere mode in the frequency domain, 
thereby explaining the variation of the delay-time distribution 
under different pulse widths and center frequencies. 
Our results further show that, 
apart from the vicinity of the primary pulse, 
the source position leaves the echo signal almost unchanged. 
The limitation of our model lies in the space-only slice of the Schwarzschild spacetime, 
leading to a different gravitational potential \cite{Resca2018}. 
The position of the photon sphere on the Flamm paraboloid 
is different from the four-dimensional case (see Appendix A for a detailed discussion). 
The extensions of this model can further take the spin of the black hole \cite{Bardeen1972,Wong2021}, wave polarizations \cite{EHT2024}, and the finite size of the light source \cite{Matsunaga2006,Zenginoglu2012,Wong2021} into account. 

This study investigated the phenomenon of echoes caused by the strong gravitational field 
on the curved surface, providing an analogue model for laboratory simulation. 
Our discussion about the interplay between the black hole's photon sphere and the source's pulse duration 
offers a new perspective on the strong gravitational lensing and lensing of transients 
and may be helpful for the detection of compact celestial objects. 
In addition, the curved waveguides have great potential for optical applications, 
such as the curved space nanostructure \cite{Bekenstein2017}, and geodesic lenses \cite{Xu2019,EFG2024}. 
The circulation of geodesics endows the system with characteristics similar to the ring resonator, 
suggesting new possibilities for the design of the optical equipment. 
The effects in curved space can lead to applications in the design of curved photonic waveguides and microcavities \cite{Song2021,Huang2024,Roth2025,Girin2025,Xu2025}. 

\acknowledgments
This work was supported by the National Natural Science Foundation of China (62375241).

\appendix
\section{The photon sphere on the Flamm paraboloid}
Photons moving on circular orbits around a compact object constitute its photon sphere. 
On the Flamm paraboloid, the geodesics also have a similar circulation behavior around $r=r_g$. 
Following the procedure in four-dimensional spacetime \cite{SeanCarroll}, 
here we calculate the position of photon sphere on the surface. 
From  Eq.~(\ref{eq:FlammMetric}) we can derive 
\begin{equation}
(\frac{dr}{ds})^2=(1-\frac{r_g}{r})(1-\frac{r_p^2}{r^2}).
\end{equation}
The meaning of the parameter $r_p$ is identical to that in Eq.~(\ref{eq:GeoOrbEq}).
The right side of the equation can be denoted as $\varepsilon -V(r)$, 
where $\varepsilon$ means the conserved effective energy and 
$V(r)$ means the effective potential. 
The photon sphere appears at the position where $dV(r)/dr=0$. 
Since the effective energy $\varepsilon$ is independent of the radial distance $r$, 
differentiating the right side of the equation leads to 
\begin{equation}
\frac{dV(r)}{dr}=-\frac{1}{r^3}\left[ r_g r \left( 1-\frac{r_p^2}{r^2} \right)+2r_p^2 \left( 1-\frac{r_g}{r}\right) \right]. 
\end{equation}
When $r=r_p$, the first term in the square brackets is zero; 
when $r=r_g$, the second term in the square brackets is zero. 
When $r=r_p=r_g$, the first derivative of the potential is zero, 
so the photon sphere occurs at $r=r_g$ on the Flamm paraboloid, which coincides with the event horizon. 
The result is consistent with the geodesic behavior. 

Since we make the assumption of constant time, 
the position of photon sphere is different from the four-dimensional case. 
For the Schwarzschild black hole, the photon sphere occurs at $r_s=3r_g/2$ 
where corresponds to the point of maximum potential energy, forming the unstable orbit \cite{SeanCarroll}. 
Based on the Fermat principle, one can derive the Fermat metric which works as another 
analogue model of the Schwarzschild black hole. 
On the 2D curved surface of the Fermat metric, 
the position of photon sphere is $\rho=3r_g/2$, 
where $\rho$ is the radial surface parameter, 
and the relationship with the radius of rotation $r$ is 
$r(\rho)=\rho/\sqrt{1-r_g/\rho}$ \cite{Perlick2004}. 

\section{The transition of the angular interference fringe number}
\begin{figure}[tbph]
\includegraphics[width=0.8\textwidth]{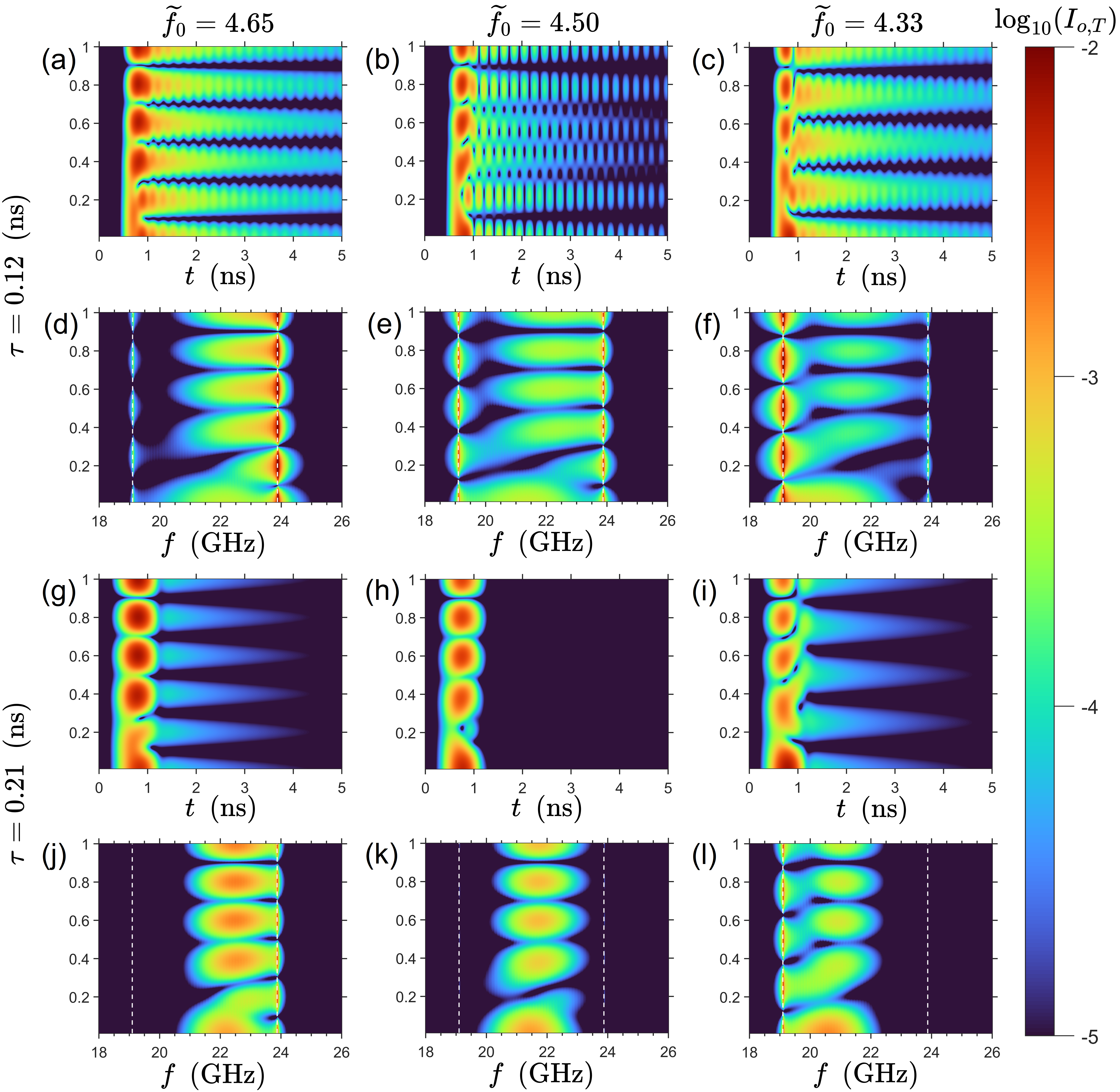}
\caption{
The temporal response and the corresponding frequency domain distribution 
under different center frequency $f_0$ and pulse half-width $\tau$. 
The center frequency $f_0$ for each column is 4.65$f_{\rm{echo}}$, 4.50$f_{\rm{echo}}$, and 4.33$f_{\rm{echo}}$. 
The first row and the third row are the temporal response under $\tau=0.12$ ns and $0.21$ ns. 
The second row and the fourth row are the corresponding frequency domain distribution. 
The white dash lines mark the position of $M=4$ photon sphere mode with $4f_{\rm{echo}}=19.10$ GHz 
and the $M=5$ photon sphere mode with $5f_{\rm{echo}}=28.37$ GHz. 
Other parameters are the same as in Fig.~\ref{fig:TimeDomain}. 
}
\label{fig:SFig_stripe}
\end{figure}
From Fig.~\ref{fig:SFig_stripe}(a) to (c), 
as the central frequency $f_0=c/\lambda_0$ decreases, 
the distribution of dark fringes changes from five to four. 
Fig.~\ref{fig:SFig_stripe}(b) is in the transition process, 
where the value of $\widetilde{f}_0=2\pi r_g/\lambda_0=4.5$ is the half-integer between five and four. 
The corresponding frequency domain distributions are shown in Fig.~\ref{fig:SFig_stripe}(d) to (f). 
In Fig.~\ref{fig:SFig_stripe}(d), the $M=5$ photon sphere mode dominates; 
in Fig.~\ref{fig:SFig_stripe}(f), the $M=4$ photon sphere mode takes over. 
While in Fig.~\ref{fig:SFig_stripe}(e), 
the $M=4$ photon sphere mode and the $M=5$ photon sphere mode are comparable. 
Therefore the corresponding echo tails interfere with each other, 
resulting in the interlaced stripes in Fig.~\ref{fig:SFig_stripe}(b). 
From Fig.~\ref{fig:SFig_stripe}(g) to (i), 
a transition from five dark fringes to four dark fringes also occurs, 
but the echo tail interference does not happen as the case of $\tau=0.12$ ns. 
During the transition process, the echo tails first disappear 
when Fig.~\ref{fig:SFig_stripe}(g) changes to Fig.~\ref{fig:SFig_stripe}(h), 
and then reappear when Fig.~\ref{fig:SFig_stripe}(h) changes to Fig.~\ref{fig:SFig_stripe}(i). 
In the frequency domain, a longer pulse width leads to a more concentrated frequency domain distribution. 
As the center frequency decreases, there exist situations 
where no photon sphere mode is excited (see Fig.~\ref{fig:SFig_stripe}(k)), 
and thus no echo tail appears in the corresponding time domain (see Fig.~\ref{fig:SFig_stripe}(h)). 
Such transition in the echo tails further confirms the view 
that each photon sphere mode in the frequency domain corresponds to a series of echo tails. 

\section{The length difference between $n=0$ and $n=1$ anti-clockwise geodesics}
\begin{table}[htbp]
\caption{
\label{tab:D_L3L1}
The relative geodesic difference $(L_3-L_1)/\lambda_0$ with 
the incident central wavelength $\lambda_0=1$ cm, $2\pi/6 \approx 1.05$ cm, and $1.5$ cm. 
The second column corresponds to Fig.~\ref{fig:TimeDomain_different_r0}(b), 
and the third column corresponds to Fig.~\ref{fig:TimeDomain_different_r0}(d). 
Other parameters are the same as in Fig.~\ref{fig:TimeDomain}. 
}
\begin{ruledtabular}
\begin{tabular}{cccc}
$r_i$ cm & $\lambda_0=1$ cm & $\lambda_0=2\pi/6$ cm & $\lambda_0=1.5$ cm \\
\colrule
1.1 & 6.4630 & 6.1717 & 4.3087 \\
8 & 7.6870 & 7.3406 & 5.1247 \\
12 & 7.8462 & 7.4926 & 5.2308 \\
16 & 7.9438 & 7.5858 & 5.2959\\
\end{tabular}
\end{ruledtabular}
\end{table}
Here, we further explain the results shown in Fig.~\ref{fig:TimeDomain_different_r0}(b) and (d) 
when $\varphi_{out}=\pi$ from the perspective of the relative length difference $(L_3-L_1)/\lambda_0$. 
From the third column of Table~\ref{tab:D_L3L1}, 
it can be seen that except for the $r_{i}=1.1$ cm case, 
all other cases possess the values of $(L_3-L_1)/\lambda_0$ close to half-integers. 
Therefore, in the output intensity curve in Fig.~\ref{fig:TimeDomain_different_r0}(b), 
the cases of $r_{i}=8$ cm, 12 cm, and 16 cm all exhibit double peaks 
due to the destructive interference between adjacent wave packets. 
Similarly, from the fourth column of Table~\ref{tab:D_L3L1}, 
when $r_i=1.1$ cm and 16 cm, the value of $(L_3-L_1)/\lambda_0$ is closer to the half-integer, 
and the corresponding light intensity curves in Fig.~\ref{fig:TimeDomain_different_r0}(d) 
also have double peaks.

\end{document}